\documentstyle[prl,aps]{revtex}
\begin{document}
\title{Asymptotically flat solutions to the Ernst equation
with reflection symmetry} 
\author{R. Meinel and G. Neugebauer}
\address{Max--Planck--Gesellschaft, 
Arbeitsgruppe Gravitationstheorie \\ an der Universit\"at 
Jena, Max-Wien-Platz 1, D-07743 Jena, Germany \\
}
\maketitle
\begin{abstract}
It is shown that the class of asymptotically flat solutions to the 
axisymmetric and
stationary vacuum Einstein equations with reflection symmetry of the 
metric is uniquely
characterized by a simple relation for the Ernst potential $f^{(u)}$ on 
the upper part 
of the
symmetry axis ($\zeta$--axis): $f^{(u)}(\zeta)\bar{f}^{(u)}(-\zeta) = 1$. 
This result generalizes 
a well--known fact from potential theory: Axisymmetric 
solutions to the Laplace equation
that vanish at infinity and have reflection symmetry with 
respect to the plane $\zeta = 0$
are characterized by a potential that is an odd function 
of $\zeta$ on the upper part
of the $\zeta$--axis.  
\end{abstract}
\section{Introduction}
The general solution of the axisymmetric Laplace equation 
\begin{equation}
\triangle U \equiv U,_{\rho\rho}+\frac{1}{\rho}U,_{\rho} + 
U,_{\zeta\zeta} = 0
\label{lap}
\end{equation}
that is regular outside some finite region and vanishes at infinity 
may be written in form of a multipole expansion
\begin{equation}
U = \sum_{n=0}^{\infty} c_n\, r^{-(n+1)}\, {\rm {P_n}}(\cos \theta)
\end{equation}
with
\begin{equation}
\rho = r\sin \theta, \quad \zeta = r\cos \theta.
\end{equation}
Solutions with reflection symmetry 
\begin{equation}
U(\rho,-\zeta)=U(\rho,\zeta) 
\label{us1}
\end{equation}
are given by
\begin{equation}
c_n = 0 \quad\mbox{for}\quad n = 1,3,5,\dots
\end{equation}
This leads to the following form of the potential $U^{(u)}$ on 
the upper part of the 
axis ($\theta = 0$):
\begin{equation}
U^{(u)}(\zeta)=\sum_{k=0}^\infty c_{2k}\, \zeta^{-(2k+1)}.
\end{equation}
Hence, solutions with reflection symmetry are uniquely characterized by an odd
function $U^{(u)}(\zeta)$:
\begin{equation}
U^{(u)}(-\zeta) = - U^{(u)}(\zeta).
\label{us2}
\end{equation}

The aim of the present paper is to prove a relation that 
generalizes Eq. (\ref{us2})
to the case of solutions to the axisymmetric and stationary 
vacuum Einstein equations with
reflection symmetry. As is well known these equations are equivalent to an
equation for the complex Ernst potential $f(\rho,\zeta)$:
\begin{equation}
(\Re f)\triangle f = (\nabla f)^2.
\label{ernst}
\end{equation}
Solutions with reflection symmetry of the metric are defined by 
\begin{equation}
f(\rho,-\zeta) = \bar{f}(\rho,\zeta),
\label{fs1}
\end{equation}
where a bar denotes complex conjugation.
Eq. (\ref{fs1}) is the analogue of Eq. (\ref{us1}).
(It should be noted that the real part $\Re f$ of $f$
is  directly related to some metric
coefficient, whereas its imaginary part is related to some 
other metric coefficient in
such a way, that symmetry of the metric means antisymmetry of $\Im f$.) 
The analogue of Eq. (\ref{us2}) will be shown to be
\begin{equation}
f^{(u)}(\zeta)\bar{f}^{(u)}(-\zeta) = 1.
\label{fs2}
\end{equation}

The Laplace equation (\ref{lap}) is a special case of the 
Ernst equation (\ref{ernst})
for real Ernst potentials 
\begin{equation}
f = \exp (2U).
\end{equation}
In this way it can easily be verified that Eqs. (\ref{us1}) 
and (\ref{us2}) are indeed special cases
of (\ref{fs1}) and (\ref{fs2}), respectively.

Our prove of relation (\ref{fs2}) is based upon the existence of a 
Linear Problem whose
integrability condition is just the Ernst equation.   
 
\section{The Linear Problem related to the Ernst equation}
Introducing complex variables 
\begin{equation}
z=\rho+i\zeta, \quad \bar{z}=\rho-i\zeta
\end{equation}
the Ernst equation takes the form
\begin{equation}
(f+\bar{f})\{(\rho f,_{\bar{z}}),_z+(\rho f,_z),_{\bar{z}}\}=
4\rho f,_zf,_{\bar{z}}.
\label{KErnst}
\end{equation}
Eq. (\ref{KErnst}) is the integrability condition of a related Linear Problem 
for the 
$2\times 2$--matrix function $\Phi(\lambda,z,\bar{z})$:
\begin{equation}
\Phi,_z=\left\{ \left( \begin{array}{cc} N & 0 \\ 0 & M \end{array} \right)
        +\lambda \left( \begin{array}{cc} 0 & N \\M & 0 \end{array} 
        \right) \right\}\Phi,
\label{Lin1}
\end{equation}
\begin{equation}
\Phi,_{\bar{z}}=\left\{ \left( \begin{array}{cc} \bar{M} & 0 \\ 0 & \bar{N} 
         \end{array} \right)
        +\frac{1}{\lambda} \left( \begin{array}{cc} 0 & \bar{M} \\ \bar{N} & 0 
        \end{array} \right) \right\}\Phi,
\label{Lin2}
\end{equation}
cf.~\cite{lp}. $M$ and $N$ depend on $z$ and $\bar{z}$ but not 
on $\lambda$, which is defined as
\begin{equation}
\lambda = \sqrt{\frac{K-i \bar{z}}{K+iz}},
\label{lam}
\end{equation}
where $K$ is an additional complex variable called the spectral 
parameter\footnote{For some considerations of the matrix function
$\Phi$ it turns out to be appropriate to switch to a different set of
independent variables, e.~g.  $K, \rho, \zeta$  instead of  $\lambda, z,
\bar{z}$.}.
Consequently,
\begin{equation}
\lambda,_z=\frac{\lambda}{4\rho}(\lambda^2-1), \quad 
              \lambda,_{\bar{z}}=\frac{1}{4\rho\lambda}(\lambda^2-1).
\end{equation}
The condition $\Phi,_{z\bar{z}}=\Phi,_{\bar{z}z}$ then implies a first order 
system of nonlinear equations
for the functions $M$ and $N$ which is equivalent to the Ernst equation via
\begin{equation}
M=\frac{f,_z}{f+\bar{f}}, \quad N=\frac{\bar{f},_z}{f+\bar{f}}.
\label{tet}
\end{equation}
Without loss of generality the following structure of the 
matrix $\Phi$ may be assumed:
\begin{equation}
\Phi=\left( \begin{array}{cr} \psi(\lambda,z,\bar{z}) & 
\psi(-\lambda,z,\bar{z}) \\ \chi(\lambda,z,\bar{z}) & -\chi(-\lambda,z,\bar{z})  
\end{array} \right),
\end{equation}
together with
\begin{equation}
\overline{\psi(\frac{1}{\bar{\lambda}},z,\bar{z})}=\chi(\lambda,z,\bar{z}).
\label{kom}
\end{equation}
For $K\rightarrow\infty$ and $\lambda = -1$ the 
functions $\psi$ and $\chi$ may be 
normalized 
by
\begin{equation}
\psi(-1,z,\bar{z}) = \chi(-1,z,\bar{z}) = 1.
\label{norm}
\end{equation}
The related
solution to the Ernst equation is given by
\begin{equation}
f(\rho,\zeta)\equiv \chi(1,z,\bar{z}) \quad (K\rightarrow\infty).
\end{equation} 
We mention that, according to (\ref{lam}), the 
complex $\lambda$--plane is related to a
two--sheeted Riemann $K$--surface, with the 
branching points $K={\rm i}\bar{z}$ and
$K=-{\rm i}z$.

\section{Solution of the Linear Problem on the axis}
We consider solutions of the Ernst equation that are regular outside some
finite region. Therefore, on the axis $\rho=0$, we distinguish two regions:
an upper part $\zeta>\zeta_+$  and a lower part $\zeta<\zeta_-$, 
where we assume
the solution to be regular. We introduce the following
notation:
\begin{equation}
f^{(u)}(\zeta) \equiv f(0,\zeta), \quad \mbox{with} \quad \zeta>\zeta_+;
\quad
f^{(l)}(\zeta) \equiv f(0,\zeta), \quad \mbox{with} \quad \zeta<\zeta_-.
\end{equation}
For $\rho=0$ and $K\neq\zeta$ Eq. (\ref{lam}) leads to 

\begin{equation}
\lambda = \pm 1.
\end{equation}
This allows us to integrate the Linear Problem  (\ref{Lin1}), (\ref{Lin2})
along the axis\footnote{Note that, for any function $\varphi$ depending on
$\lambda, z, \bar{z}$, $\lim_{\rho\to 0} \varphi(\lambda,z,\bar{z})$ is a
function of the {\it two} variables $K$ and $\zeta$, in general. For
example, $\lim_{\rho\to 0} (1-\lambda^2)/(z+\bar{z}) =  \lim_{\rho\to 0}
i/(K+iz) = i/(K-\zeta)$.}:
\begin{equation}
\left( \begin{array}{c}\psi\\ \chi\end{array}\right) = 
F_{\pm}^{(u/l)}(K) \left( \begin{array}{r}\bar{f}^{(u/l)}(\zeta)\\ 
\pm f^{(u/l)}(\zeta)\end{array} \right)
+ G_{\pm}^{(u/l)}(K) \left( \begin{array}{r}1\\ \mp 1\end{array}\right).
\label{achs}
\end{equation}
Two of these 8 integration `constants' $F_{\pm}^{(u/l)}(K)$ 
and $G_{\pm}^{(u/l)}(K)$ 
(they may depend on $K$, but not on $\zeta$) can be fixed by 
prescribing $\psi$
and $\chi$ at one point $\zeta=\zeta_0$ on 
the axis, for $K$--values from one of the
two sheets of the Riemann surface ($\lambda=+1$ or $\lambda=-1$). 
We choose some $\zeta_0>\zeta_+$ and assume for $\lambda=-1$
\begin{equation}
\rho=0, \zeta=\zeta_0, \lambda=-1: \quad \psi \equiv 1, 
\quad \chi \equiv 1.
\label{fix}
\end{equation} 
This choice is consistent with Eqs.~(\ref{kom}) and (\ref{norm}). 
From Eqs.~(\ref{achs}) and (\ref{fix}) we conclude
\begin{equation}
F_-^{(u)}(K)\equiv 0, \quad G_-^{(u)}(K)\equiv 1.
\label{fix1}
\end{equation}  
Now we use the assumption of asymptotic flatness, i.~e.
\begin{equation}
f\rightarrow 1 \quad \mbox{as}\quad \rho^2+\zeta^2\rightarrow\infty.
\end{equation}
As a consequence, the coefficients $M$ and $N$ in 
the Linear Problem (\ref{Lin1}),
(\ref{Lin2})
vanish asymptotically such that $\psi$ and $\chi$ do not change on the 
half--circle in the $\rho$--$\zeta$--plane
\begin{equation}
\rho=R \sin\theta,\quad \zeta=R \cos \theta, \quad 0\le\theta\le\pi,
\end{equation}
in the limit $R\rightarrow\infty$.
In this limit we find from (\ref{lam})
\begin{equation}
\lambda = \pm \exp({\rm i}\theta),
\end{equation}
i.~e. $\lambda$ changes from $\pm 1$ to $\mp 1$ as $\theta$ changes 
from $0$ to $\pi$.
Therefore, we may conclude from Eqs.~(\ref{achs}) 
for $\zeta\rightarrow\pm\infty$
\begin{equation}
F_{\pm}^{(u)} + G_{\pm}^{(u)} = F_{\mp}^{(l)} + G_{\mp}^{(l)}, \quad
F_{\pm}^{(u)} - G_{\pm}^{(u)} = -F_{\mp}^{(l)} + G_{\mp}^{(l)},
\end{equation}
i.~e.
\begin{equation}
F_{\pm}^{(u)} = G_{\mp}^{(l)}, \quad G_{\pm}^{(u)} = F_{\mp}^{(l)}.
\end{equation}
This means, together with Eq.~(\ref{fix1}), that all these functions may be
expressed in terms of two of them,
say $F_{+}^{(u)}$ and $G_{+}^{(u)}$. 
With the new notation
\begin{equation}
F(K)\equiv F_{+}^{(u)}(K), \quad G(K)\equiv G_{+}^{(u)}(K)
\end{equation}
we may summarize these findings as follows
\begin{equation}
\begin{array}{lcl}\underline{\zeta>\zeta_+:}\quad & \lambda=-1:\quad 
      & \psi=\chi=1,\\
      & \lambda=+1: \quad& \psi=F(K)\bar{f}^{(u)}(\zeta)+G(K), 
                           \quad \chi=F(K)f^{(u)}(\zeta)-G(K),\\
\underline{\zeta<\zeta_-:}\quad & \lambda=+1:\quad & \psi=\bar{f}^{(l)}(\zeta),     
        \quad \chi=f^{(l)}(\zeta),\\
      & \lambda=-1: \quad& \psi=G(K)\bar{f}^{(l)}(\zeta)+F(K),
       \quad \chi=-G(K)f^{(l)}(\zeta)+F(K). 
\end{array}                                       
\label{resa}
\end{equation}
The functions $F(K)$ and $G(K)$ characterize a solution $f(\rho,\zeta)$ of the
Ernst equation uniquely. They are directly related 
to the Ernst potential on the axis.
This can be seen by taking $K=\zeta$, i.~e. by 
going to the branching points of
Eq. (\ref{lam}). (Note that $K={\rm i}\bar{z}$ and $K=-{\rm i}z$ coincide 
for $\rho=0$.) 
For $K=\zeta$ the values of $\psi$ and $\chi$ must 
be unique, i.~e. the formulae
for $\lambda=+1$ and $\lambda=-1$ in Eqs. (\ref{resa}) have to be identified.
This leads to
\begin{equation}
f^{(u)}(\zeta)=\frac{1+G(\zeta)}{F(\zeta)},\quad 
  \bar{f}^{(u)}(\zeta)=\frac{1-G(\zeta)}{F(\zeta)},\quad
   f^{(l)}(\zeta)=\frac{F(\zeta)}{1+G(\zeta)},\quad
   \bar{f}^{(l)}(\zeta)=\frac{F(\zeta)}{1-G(\zeta)}.
\label{rel}
\end{equation}
On the other hand, we obtain
\begin{equation}
F(\zeta)=\frac{1}{\Re f^{(u)}(\zeta)}, \quad 
G(\zeta)={\rm i}\,\frac{\Im f^{(u)}(\zeta)}{\Re f^{(u)}(\zeta)}.
\label{FG}
\end{equation}
We note that the function $F(K)$ is related to the determinant 
of $\Phi(\lambda,z,\bar{z})$.
From Eqs.~(\ref{Lin1}), (\ref{Lin2}) and (\ref{tet})
one finds that $(\Re f)^{-1}\,{\rm det}\,\Phi(\lambda,z,\bar{z})$ does 
not depend on $z$ 
and $\bar{z}$. Calculating \,${\rm det}\,\Phi$ \,for\, $\rho=0$, 
$\zeta\rightarrow\infty$\, then
leads to the result
\begin{equation}
F(K)=-\frac{1}{f+\bar{f}}\,{\rm det}\,\Phi(\lambda,z,\bar{z}).
\end{equation} 

\section{Solutions with reflection symmetry}
The symmetry relation (\ref{fs1}) reads on the axis
\begin{equation}
f^{(l)}(-\zeta)=\bar{f}^{(u)}(\zeta) \quad (\zeta>\zeta_+>0).
\label{beza}
\end{equation}
(Because of the symmetry we may assume $\zeta_-=-\zeta_+$.)
Then we find from Eqs.~(\ref{rel}) 
\begin{equation}
G(-\zeta)=G(\zeta), \quad G^2(\zeta)=1-F(\zeta)F(-\zeta).
\label{gf}
\end{equation}
These relations together with (\ref{FG}) lead to
\begin{equation}
\Re f^{(u)}(\zeta)\, \Im f^{(u)}(-\zeta) = \Re f^{(u)}(-\zeta)\, 
\Im f^{(u)}(\zeta),
\quad \Re f^{(u)}(\zeta)\, \Re f^{(u)}(-\zeta) + \Im f^{(u)}(\zeta)\, 
\Im f^{(u)}(-\zeta)=1.
\label{reim}
\end{equation}
This is equivalent to relation (\ref{fs2}) that was to be shown. 
If, on the other hand,
Eq.~(\ref{fs2}) is satisfied, one can easily invert the 
above conclusions to show 
Eq.~(\ref{beza}), and from this, by the symmetric 
structure of the Ernst equation itself, the
symmetry  (\ref{fs1}) in the whole space may be concluded. 
Hence we are lead to the
following statement:

{\it An asymptotically flat solution to the Ernst equation 
has reflection symmetry (\ref{fs1})
if and only if the relation (\ref{fs2}) is satisfied.}  

It is interesting to note that the reflection 
symmetry (\ref{fs1}) of the Ernst potential leads 
to the following global property of $\Phi(\lambda,z,\bar{z})$:
\begin{equation}
\frac{1}{f+\bar{f}}\,\Phi^T(\lambda,z,\bar{z})\,
\Phi(-\frac{1}{\lambda},\bar{z},z) =
\left(\begin{array}{lr}1&G(K)\\G(K)&1\end{array}\right)
\end{equation}
together with
\begin{equation}
G(-K)=G(K),\quad G^2(K)=1-F(K)F(-K).
\end{equation}
$F(K)$ and $G(K)$ are analytic functions, i.~e. they 
may be obtained as analytic continuations
of $F(\zeta)$ and $G(\zeta)$ defined by means of the axis 
potential according to (\ref{FG}).

\section{Discussion}
Within the Newtonian theory of gravitation all equilibrium  states of isolated
self--gravitating fluids (i.~e. stellar models) must have reflection symmetry
through a plane (the `equatorial plane') which is perpendicular to the
rotation axis of the star, see \cite{lich}. It has been conjectured that 
stationary general relativistic stellar models must have reflection
symmetry as well \cite{l}. Therefore, reflection symmetric solutions to the 
Ernst equation describing the exterior gravitational fields of rotating
bodies are of particular interest. Also the 
Kerr solution describing a rotating
black hole has reflection symmetry. Accordingly, the Ernst potential of the
Kerr solution satisfies (\ref{fs2}). The same holds true for the 
solution describing
a rigidly rotating disk of dust \cite{NM1}, \cite{NM2}.
A consequence of relation (\ref{fs2}) is, that -- as in the case of the
Laplace equation -- every second multipole moment vanishes. The 
non--vanishing  moments are $M_0, M_2, M_4, \dots$ for the `mass moments'
and $J_1, J_3, J_5, \dots$ for the `rotational moments'. (For a definition
of these moments see \cite{mom}.) This simplifies the investigation of
reflection symmetric solutions. 
 
\section*{Acknowledgement}
We acknowledge an interesting discussion with Panagiotis Kordas who intends to
publish an independent  proof of  
relation (\ref{fs2})
 \cite{PK}. In fact, this 
discussion stimulated us to publish our result, found as a by--product of our
work on the rigidly rotating 
disk of dust \cite{NM1}, \cite{NM2}. The relations 
(\ref{reim}) were necessary for deriving Eq.~(24) in \cite{NM1}.

\end{document}